\documentclass[11pt]{article}

\def\fnote#1#2{\begingroup\def\thefootnote{#1}\footnote{#2}\addtocounter
{footnote}{-1}\endgroup}

\usepackage{amsmath, amsthm, amssymb,slashed,graphicx,epsfig}

\newcommand {\slsh} [1] {\not{\hbox{\kern-2pt${#1}$}}}

\newcommand{\gsim}{\lower.7ex\hbox{$\;\stackrel{\textstyle>}{\sim}\;$}}
\newcommand{\lsim}{\lower.7ex\hbox{$\;\stackrel{\textstyle<}{\sim}\;$}}
\newcommand {\beq} {\begin{equation}}
\newcommand {\eeq} {\end{equation}}
\newcommand {\beqn}{\begin{eqnarray}}
\newcommand {\eeqn} {\end{eqnarray}}
\newcommand{\bea}{\begin{eqnarray}}
\newcommand{\eea}{\end{eqnarray}}

\def\tg{\widetilde{g}}
\def\p{\varphi}
\def\tp{\widetilde{\varphi}}

\def\U{{\rm U}}

\def\1{\mathbbm{1}}
\def\N{{\cal N}}

\def\p{{}^{\,\prime}}

\def\MP{M_{\rm P}}
\def\LP{L_{\rm P}}
\def\p{\varphi}
\def\GN{G_{\rm N}}

\def\tGN{\widetilde{G}_{\rm N}}

\def\tnabla{\widetilde{\nabla}}
\def\tf{\widetilde{f}}
\def\S{\mathbb{S}}

\begin{document}

\begin{titlepage}
\begin{flushright}
FTPI-MINN-09/27; UMN-TH-2807/09
\\
08/20/09
\end{flushright}
 \vspace{56pt}

\begin{center}
{\large {\bf {On the Possibility of a Trans-Planckian Duality}}}

\vspace{36pt}
Stefano Bolognesi\fnote{*}{ 
bolognesi@physics.umn.edu}\\
\vspace{20pt}
{\em  William I. Fine Theoretical Physics Institute, University of Minnesota, \\ 116 Church St. S.E., Minneapolis, MN 55455, USA.}

\vspace{30pt}

\noindent
{\bf Abstract}
\end{center}
\noindent 
We investigate the possibility of a trans-Planckian duality, which exchanges a manifold of events (space-time), with a manifold of momenta (energy-momentum). 
Gravity has a dual counter-part, that is, a geometric theory defined on the  manifold of momenta.
We provide a mathematical framework that can possibly realize this idea, and analyze its classical behaviour.

\vfill

\end{titlepage}

\tableofcontents

\section{Introduction}

The notorious problem of gravity is that, treated as a quantum field theory in the perturbative expansion, it is not renormalizable.
It makes sense, though,  to consider it as an effective field theory, at a certain energy scale $\Lambda$. 
When the energy scale is much lower than the Planck mass $\MP$, gravity is in the semi-classical regime, and thus follows the rules of general relativity, plus tiny quantum corrections.
As we increase the energy scale, the effective coupling constant of gravity becomes stronger and stronger. This is simply a consequence of the fact that geometry couples directly to the energy-momentum tensor, and the strength of the coupling is essentially measured by the ratio $\Lambda/\MP$.
In some sense, gravity at high energy, is similar to the problems that we are used to encountering in the infrared behaviors of asymptotically free quantum field theories. The problem of quantum gravity thus can also be considered  as a strong coupling problem, not in the infrared (IR), but in the ultraviolet (UV).

One of the most fascinating ideas, which has emerged in the study of strongly coupled quantum field theories, is that of duality.\cite{Coleman:1974bu}
By now, we know a broad variety of dualities, each one with its own peculiarity. 
They all, though, share a common feature. 
A theory which becomes intractable in certain regimes (energy scales or coupling constants) can be described by another theory, a dual one, which has instead a semi-classical, simpler description. 
These dualities generally exchange strong coupling with weak coupling. 
Dual theories are generally written in terms of different variables, bound states or solitonic objects of the original theory.
Sometimes, although written in terms of different variables, the two theories are identical, in which case we call them self-dual.

We mention a few examples.  
One realized in nature, in a broad sense of duality, is ordinary QCD. At high energy scales, QCD is well described by a gauge theory with a certain  number of colors and flavors. The gauge couplings becomes strong in the IR, and every perturbative computation, performed with the original Lagrangian, becomes unreliable. 
There is though another description, at energies much lower than the dynamical scales. 
We can write an effective Lagrangian in terms of pions, bound states of quarks, which are the Goldstone boson of the chiral symmetry breaking.  One of the first, exact theoretical realizations of the strong-weak duality is probably sine-Gordon/Thirring  correspondence in $1+1$ dimensions. 
Supersymmetry made it possible to find concrete realizations in $3+1$ theories, in particular the Seiberg-Witten solution in $\N=2$, and the Olive-Montonen duality of $\N=4$ Super-Yang-Mills, which is a perfect realization of the electro-magnetic duality.

Gravity becomes strong at the Planck scale, and it becomes ultra-strong if we increase the energy even further, in the trans-Planckian region $\Lambda \gg \MP$. 
The purpose of this paper is to answer the following question: {\it If there is a dual description of gravity at the trans-Planckian scales, what it should look like?}

It is good to start with electro-magnetism. 
The concept of electro-magnetic duality is an old one, at least as old as the Maxwell equations. 
It is very simple to explain the concept of electro-magnetic and strong-weak duality; it is just the exchange of the electric and magnetic variables.
It is by far more complicated to find a theory that explicitly realize these concepts. 
In other words, electro-magnetism contains from the beginning the duality between electric and magnetic fields. Less trivial is to find a theory in which the sources, in some way,  also present this duality.
For gravity instead, is not even clear how to define, an equivalent concept of duality.  
The scope of this paper is to define something analogous to electro-magnetic, or strong-weak, duality for the gravitational field.

\begin{figure}[h!t]
\epsfxsize=10cm
\centerline{\epsfbox{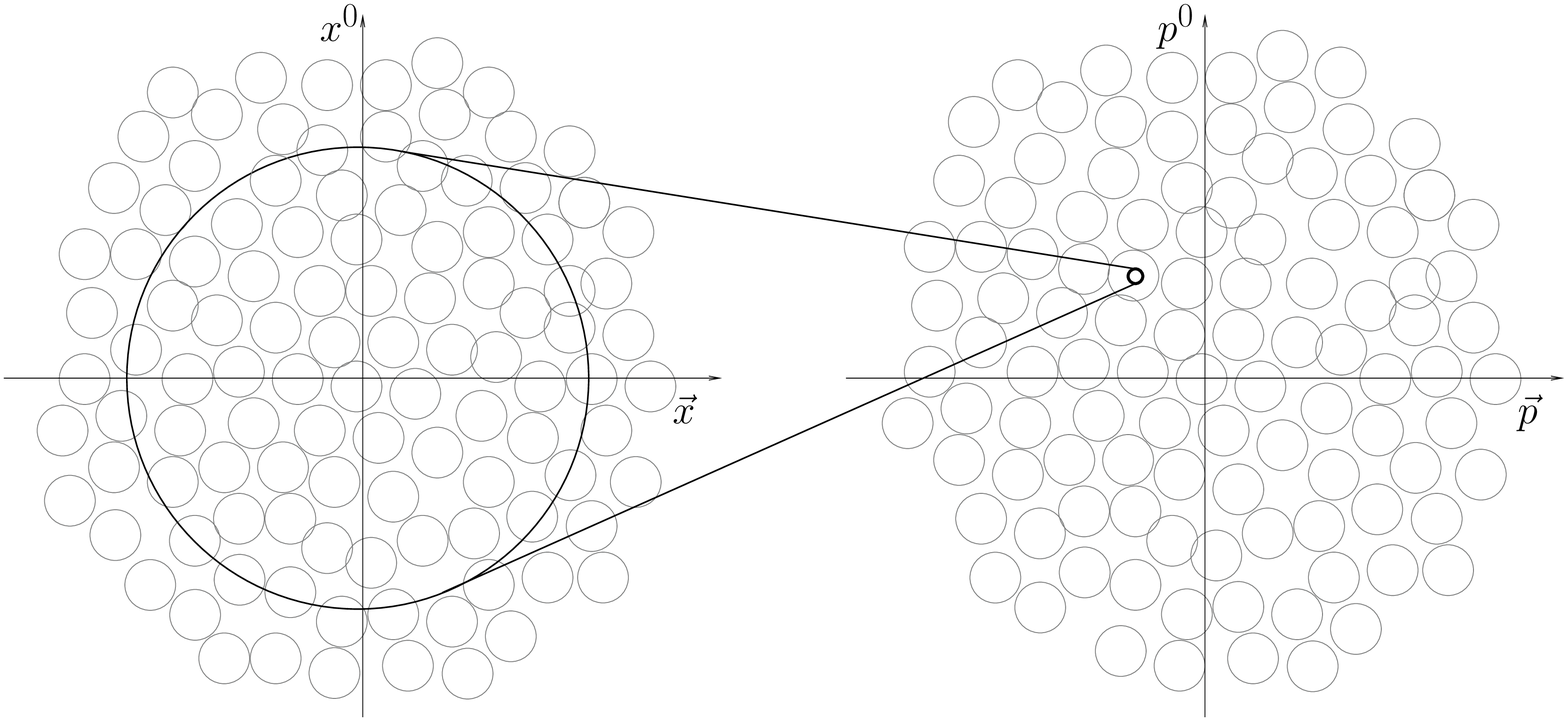}}
\caption{{\footnotesize We certainly observe a very asymmetric relation between events and momenta.
This may be due to the limited energies we can access.
The fundamental theory, may instead be completely symmetric. }}
\label{asymmetry}
\end{figure}

The universe, as we are able to observe, looks completely asymmetrical between space-time and energy-momentum. 
We could wonder if the asymmetry of our equations is just a consequence of the asymmetric conditions in which we live, and not a fundamental property of nature.
We know that gravity, through the Newton constant, introduces a new fundamental quantity. 
This quantity sets the natural unit of space-time, the Planck length $\LP$, and the natural unit of energy-momentum, the Planck mass $\MP$. 
In all the physical situations which we are able to explore, we always have to deal with space-time scales much bigger than the fundamental scale $\LP$, and energy-momentum scales much smaller than the fundamental scale $\MP$ (we can visualize this concept with the help of Figure \ref{asymmetry}).
The asymmetry of our equations could just be the consequence of the asymmetry of our environmental conditions, and not a property of the fundamental theory. 

In this paper we want to make an attempt to write a theory in which space-time and energy-momentum enter in an equivalent way.
The reason that brought us to this quest is very simple. The strength of the gravitational coupling depends on the energy scale, and is proportional to $  \Lambda/\MP$; it is weak at low energy, strong at high energy. 
If we want a strong-weak duality, the coupling of the dual theory must be proportional to $  \MP/\Lambda$. 
If furthermore we want this dual theory to be of the same form as the original one, i.e. a geometro-dynamics, the manifold on which it is defined must be of momenta (with dimension of energy), and not of positions (with dimension of length).

In Section \ref{kin} we discuss the preliminary duality between coordinates and momenta, without any action. 
We use as matter field a single complex scalar.
In Section \ref{dyn} we introduce the action for the matter field, 
explain what in the first place breakes the duality which was present at the kinematic level, and then add a new term to the action in order to repristinate it. 
In Section \ref{gra} we introduce the gravitational interaction, and also its dual counterpart.
In Section \ref{cla} we discuss the classical solutions to the self-dual action, both with and without gravity.
In Section \ref{qua} we briefly comment on the quantum behaviour.
In Section \ref{con} we present some conclusions.

We use convections where the signature of the Minkowski metric is $(+,-,-,-)$.
Throughout the paper, except for Appendix \ref{app}, we use natural units where $c = \hbar =1$.

\label{int}

\section{Kinematic}

Without dynamics, space-time and energy-momentum are completely symmetrical objects.  
They are both four-dimensional spaces
\beq
x^{\mu}=(x^0,\vec{x}) \qquad  p^{\mu}=(p^0, \vec{p}) \ ,
\eeq
equipped with a Minkowskian metric $\eta_{\mu \nu}$.
Fields can be written as function over space-time, or equivalently as functions over energy-momentum. 
The two formulations are related by a Fourier transformation\footnote{The Planck mass normalization is needed in order to have duality of the dimensions. $\p$ has dimension of mass, while $\widetilde{\p}$ of the inverse of a mass.}
\bea
\label{fourier}
\widetilde{\p}(p) &=&  \MP^2   \int \frac{d^4 x}{(2 \pi )^{2} } \ e^{ - i p^{\mu} x_{\mu}  } \ \p(x) \ , \nonumber \\
\p(x) &=& \frac{1}{\MP^2} \int \frac{d^4 p}{(2 \pi )^{2} } \ e^{ \ i p^{\mu} x_{\mu}  } \ \widetilde{\p}(p) \ .
\eea
One of the basic properties of this transformation, is the equivalence of the norms
\beq
\MP^2 \int d^4 x \  |\p(x)|^2 = \frac{1}{\MP^2} \int d^4 p \ |\widetilde{\p}(p)|^2 \ .
\eeq
Both in space-time and energy-momentum, the Lorentzian symmetries (rotations and boosts) can be written as the following  operator acting on the function $\p$ or on the dual function $\widetilde{\p}$:
\bea
J^{\mu\nu}  &=&  - i    \epsilon^{\mu\nu\rho\sigma} \  x_{\rho}  \frac{\partial}{\partial x^{\sigma}}  \nonumber \\
&=&  - i \epsilon^{\mu\nu\rho\sigma} \  p_{\rho}  \frac{\partial}{\partial p^{\sigma}}  \ .
\label{generators}
\eea
Note the self-duality of the Lorentz symmetries.

Self-duality, instead, is not a property of the translation invariance. Translation in the space-time is generated by $P^{\mu}$, while translation in the energy-momentum is generated by $X^{\mu}$:
\beq
P^{\mu}  = \  - i \frac{\partial}{\partial x_{\mu}} \ , \qquad X^{\mu}   =   i \frac{\partial}{\partial p_{\mu}} \ .
\eeq
These two transformations are not dual to each other.

Finally, in the scaling transformation, if space-time is scaled by a factor $\lambda$, energy-momentum must be scaled by the inverse factor: 
\beq
x \to \lambda x \ , \qquad p \to \lambda^{-1} p \ .
\eeq
Note that the renormalization group (RG) flow, can be obtained on both sides of the duality with this transformation. The only difference is the exchange of IR  with  UV. When the space-time theory is in the UV, at high energy scales $\Lambda \gg \MP$, the dual theory is in the IR at small length scales $1/\Lambda \ll \LP$.

\label{kin}

\section{Dynamics}

Let us now introduce the dynamics.
As we are accustomed to, we write the dynamics using the Poincare symmetry (translation plus Lorentz invariance) as our guiding principle. 
The action  of free massless particles is
\beq
\label{freescalar}
S =  \int d^4 x \ \partial_{\mu} \p^* \partial^{\mu} \p \ .
\eeq
We now see that space-time and energy-momentum enter in two completely different ways. 
This is a consequence of the mismatch between the translational invariances, as previously discussed.

\begin{figure}[h!t]
\epsfxsize=9cm
\centerline{\epsfbox{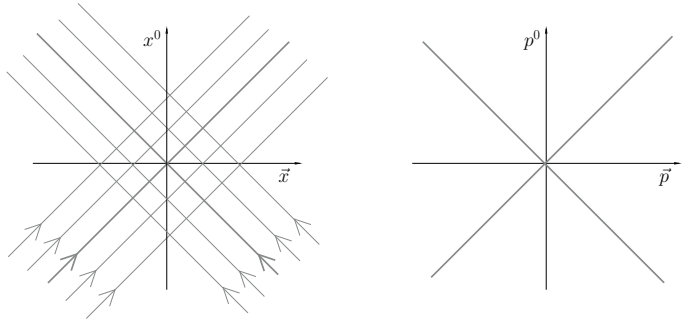}}
\caption{{\footnotesize Dynamics of free massless particles in the $x$ and $p$ spaces.}}
\label{dynamics}
\end{figure}

Solutions to the free particle action are trajectories on light-cones. 
Every trajectory corresponds to a specific point in the energy-momentum manifold, which is conserved. 
The only possible occupation points, in energy-momentum, are the ones on the light cone centered in zero (Figure \ref{dynamics}). 
Positive energies correspond to particles, negative energies correspond to anti-particles.

We can nevertheless write a ``dual'' theory, for the field $\widetilde{\p}$ defined in (\ref{fourier}):
\beq
\label{freescalardual}
\widetilde{S}  =   \int d^4 p \ \widetilde{\partial}_{\mu} \widetilde{\p}^* \widetilde{\partial}^{\mu} \widetilde{\p} \ .
\eeq
Here we used, as our guiding principles, the Poincare symmetry in the energy-momentum manifold.
Clearly the solutions of the actions $S$ and $\widetilde{S}$ are not dual to each other; i.e. they do not commute with the Fourier transformation.

As far as free particles are concerned, we can still think about the problem in a different way that makes evident a certain {\it residual} symmetric relation between space-time and energy-momentum. 
Instead of considering a particle, with a determined value of energy and momentum, traveling along space-time, we can consider a certain event in space-time, and then associate to it all the energy-momentum values that a particle or an anti-particle passing through this event, could have.
We thus recover an entire trajectory in the light cone of the energy-momentum manifold, as if it were space-time that propagates in the energy-momentum manifold and not vice-versa. 
A dual-trajectory, or a dual worldline, corresponds to many different trajectories in the ordinary $x^{\mu}$ space. We will return to this idea in Section \ref{cla}.

The previous perspective, can work only for a ``centered'' space-time, i.e. for the light-cone of trajectories passing through a certain fixed point, $x^{\mu} =0$.
But we know that the solutions of the action $S$ also include all the other light-cones in space-time, centered in any possible event $x^{\mu}$ (see Figure \ref{dynamics}).
It is thus important to write the original theory, the action $S$, in a way that also makes it possible to translate in the $p^{\mu}$ space. 
In the action (\ref{freescalar}),  $p^{\mu}=0$ is the center of the energy-momentum, and clearly has a privileged role. 
What we want is to make the theory independent upon the choice of this center.
We can do this in the following way. We replace the derivative by a covariant derivative 
\beq
\label{covariant}
\nabla_{\mu} = \partial_{\mu} -i Q_{\mu} \ ,
\eeq 
where $Q_{\mu}$ is an auxiliary, non-dynamical, gauge boson.
A translation $Q \to Q + \delta$ corresponds to the transformation $\p \to e^{i \delta x } \p$.
The gauge boson $Q$ allows us to center at any point of the energy-momentum manifold. 
We thus have built into our action $S$ the possibility to translate in the energy-momentum manifold.
The $x \leftrightarrow p$ duality and the translational invariance, require the introduction of this additional structure: a $\U(1)$ gauge bundle on both spaces, with $Q_{\mu}$ the gauge boson on the $x^{\mu}$, and $Y_{\mu}$ the gauge boson on the $p^{\mu}$. In order to implement gauge invariance, the Fourier transform will also be modified (see next Section).

The most trivial thing we can do to make an action invariant is to take the sum of two, the one written over $x$ (\ref{freescalar}), and the one written over $p$ (\ref{freescalardual}): 
\beq
\label{sum}
\S =   S  +     \widetilde{S} \ .
\eeq
This is basically the idea behind the whole paper. 
The reason behind the choice of the  plus sign shall be explained later.

We want to compare the magnitude of the two actions $S$ and $\widetilde{S}$ at a certain energy scale $\Lambda$, and see which one is dominant.
Consider the variation of the field $\p$ which is different from zero inside a four-volume $\Lambda^{-4}$, and vanishes outside. Call $\langle \p \rangle$ the average value inside. A quick estimate gives $S \sim \Lambda^{-2} {\langle \p \rangle}^2$.  From the basic properties of the Fourier transform, we know that $\widetilde{\p}$ is spread over a dual four-volume $\Lambda^4$, and the typical value is $\langle \widetilde{\p} \rangle \sim \langle \p \rangle \Lambda^{-4} \MP$. Another quick estimate gives $\widetilde{S} \sim  \Lambda^{2} \langle \widetilde{\p} \rangle^2 \sim  \Lambda^{-6} \langle \p \rangle^2 \MP^4$.
The outcome is that, at energy scales $\Lambda \ll \MP$, the dual action $\widetilde{S}$ by far dominates over the $S$. We will return to this point in Section \ref{cla} when we discuss the classical solutions.

\label{dyn}

\section{Gravity}

Dynamics, as we saw, completely spoils the events$\leftrightarrow$momenta duality.
But if we define a new action $\S$  as the sum of the two actions $S$ and $\widetilde{S}$, we have a theory which is, by definition, self-dual.
We want to show here how to include gravity into this setup.

Gravity, when added only in the action $S$, makes the asymmetry between space-time and energy-momentum even worse.
Space-time becomes a geometric manifold, and a dynamical object. 
The metric is allowed to depend on the position, and the Einstein equation tells us how the geometry is dynamically related to the energy-momentum tensor. 
The action $S$, with the introduction of gravity, is
\beq
\label{gravity}
S =  \int d^4 x \sqrt{-g} \ \left(  \frac{1}{16 \pi \GN} R   +   g^{\mu \nu} (\nabla_{\mu} \p)^* \nabla_{\nu} \p   \right) \ .
\eeq
The covariant derivative $\nabla$, as in (\ref{covariant}), contains the gauge field $Q$. 
$\GN$ is the Newton constant, and it defines the Planck mass by the relation $\MP^2 = 1/ \GN$.

The trans-Planckian action $\widetilde{S}$ is quite easy to write. 
We just have to introduce the metric $\widetilde{g}_{\mu\nu}$ of the energy-momentum into the game.
The former also becomes a dynamical object; its geometry is modified by the presence of matter. The action is:
\beq
\label{gravitydual}
\widetilde{S} =  \int d^4 p \sqrt{-\widetilde{g}} \ \left( \frac{1}{16 \pi \tGN } \widetilde{R} + \widetilde{g}^{\mu \nu} (\tnabla_{\mu} \widetilde{\p})^* \tnabla_{\nu} \widetilde{\p}   \right) \ .
\eeq
The covariant derivative $\tnabla$ also contains a gauge field $Y$:
$\tnabla_{\mu} = \partial_{\mu} -i Y_{\mu}$. 
$\tGN$ is the dual Newton constant defined by $\MP^2 = \tGN$.

If we want to realize this duality, we are forced to take both geometries into account. 
There is in fact no way to relate the geometry $g$ and $\widetilde{g}$ through the Fourier transform. 
They are {\it not} the Fourier transform of each other. 
That means that both geometries (\ref{gravity}) and (\ref{gravitydual}) will play a role in what will come. 
Note that there is here a crucial difference between the matter fields $\p$ and the metrics.
$\p$ and $\widetilde{\p}$ are the same degree of freedom, just written in a different form. The two geometries $g_{\mu \nu}$ and $\tg_{\mu\nu}$ are instead two genuinely different degrees of freedom.

\begin{figure}[h!t]
\epsfxsize=9cm
\centerline{\epsfbox{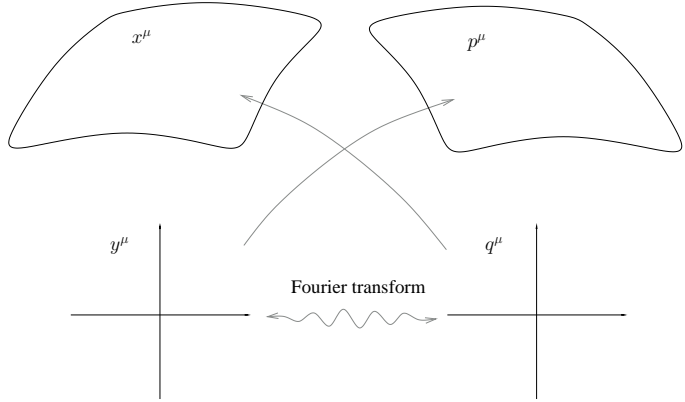}}
\caption{{\footnotesize  In order to implement the trans-Planckian duality, without ruining the equivalence principle, we need two auxiliary Minkowski spaces $y$ and $q$.}}
\label{equivalence}
\end{figure}

The challenge we now encounter, is how to relate the field $\p(x)$ and $\tp(p)$.
We want to implement the duality {\it without} ruining the equivalence and covariance principles. 
We want them to be realized both in the space-time $x$, and energy-momentum $p$ manifolds.
To implement them, we need to introduce two, auxiliary, flat Minkowski spaces, which we denote as $y$ and $q$. 
The fields $\p(y)$ and $\widetilde{\p}(q)$, are related by the ordinary Fourier transformation (\ref{fourier}).
To obtain the fields $\p(x)$ and $\widetilde{\p}(p)$, we use the following transformations (Figure \ref{equivalence}):
\bea
\label{covfourieruno}
\widetilde{\p}(p) = \MP^2 \int \frac{d^4 y} {(2 \pi )^{2} } \ \tf_{y}(p) \p(y)\ \ , \\
\label{covfourierdue}
\p(x) = \frac{1}{\MP^2} \int \frac{d^4 q} {(2 \pi )^{2} } \ f_{q}(x) \widetilde{\p}(q)\ \ , 
\eea
where functions $f_{y}(p)$ and $f_{q}(x)$, are solutions to the equation of motions (\ref{eomf}) and are asymptotically $\sim e^{-i p y}$ and $\sim e^{i q x}$ respectively. 

The space $p$, and its metric $\widetilde{g}$ , are totally unaffected by the metric $g$, as long as the boundary conditions are flat.  
It is clearly crucial, for this duality to be implementable, that the two manifolds $x$ and $p$ be asymptotically flat.

We need to consider the functions $f_q(x)$, that are the solutions of the equation of motion that asymptotically are $e^{i q x}$, with $q^2$ generic (not necessarily on-shell). The metric $g$ is asymptotically flat.
The equation of motion that $f_q(x)$ obeys, is
\beq
\label{eomf}
g^{\mu \nu} \nabla_{\mu} \nabla_{\nu} f_q(x) + \frac{\partial_{\nu} (\sqrt{-g} g^{\mu \nu})}{\sqrt{-g}} \nabla_{\mu} f_q(x)  +  (q_{\mu}  - \langle Q_\mu \rangle_{\infty})^2 f_q(x)  =  0 \ ,
\eeq
that of a scalar field, minimally coupled to gravity, with mass $q^2$, and charged under $Q_{\mu}$.  
In order for  $f_q(x)$ to be  asymptotically  $e^{i q x}$, we have to add the  terms with the asymptotic value $\langle Q_\mu \rangle_{\infty}$.
The functions $\widetilde{f}_y(p)$ satisfy the dual version of (\ref{eomf})
\beq
\label{eomfdual}
\tg^{\mu \nu} \tnabla_{\mu} \tnabla_{\nu} \tf_y(p) + \frac{\widetilde{\partial}_{\nu} (\sqrt{-\tg} \tg^{\mu \nu})}{\sqrt{-\tg}} \tnabla_{\mu} \tf_y(p)  +  (y_{\mu}  - \langle Y_\mu \rangle_{\infty})^2 \tf_y(p)  =  0 \ ,
\eeq
This is the trick that allows us to define a Fourier map between $x$ and $p$, which is consistent with the covariance and equivalence principles (and also gauge invariance) on both manifolds.

We finally define our action $\S$, which is the sum of (\ref{gravity}) and (\ref{gravitydual}), where $g_{\mu \nu}$ and $\widetilde{g}_{\mu\nu}$ are two independent asymptotically flat metrics, and the fields $\p(x)$ and $\tp(p)$ are related by the generalized Fourier maps (\ref{covfourieruno}) and (\ref{covfourierdue}).
\label{gra}

\section{Classical Theory}

We now want to study the solutions of the self-dual action $\S$, at least in the classical case. 
We begin with the theory without gravity. 
The action can be recast in a local form in space-time. Solutions to the equation of motion (EoM) are expressible through the harmonic oscillator polynomials.
We then add interactions, and make an interpretation of the result (how the theory behaves at large and small energy scales).
Finally, we discuss the gravitational case, and how the dual-geometry is perceived in space-time as a non-local interaction.

\subsection*{Classical solution without gravity}

The action is
\beq
\S =  \int d^4 x \ (\nabla_{\mu}  \p)^* \nabla^{\mu} \p  +  \int d^4 p \ (\tnabla_{\mu} \widetilde{\p})^* \tnabla^{\mu} \widetilde{\p} \ , 
\eeq 
with
\beq
\nabla_{\mu} = \frac{\partial}{\partial x^{\mu}} -i Q_{\mu} \ , \qquad 
\tnabla_{\mu}=\frac{\partial}{\partial p^{\mu}} -i Y_{\mu} \ ,
\eeq 
where $Q_{\mu}$ and $Y_{\mu}$ are the auxiliary  gauge bosons.
We rewrite it as a function of $\p$, using the inversion of the Fourier transform (\ref{fourier}):
\beq
\S =  \int d^4 x \ \Big[ |(\partial_{\mu}-iQ_{\mu})  \p|^2 
+ ( \MP)^4 (x_{\mu}-Y_{\mu})^2  |\p|^2  \Big] \ .
\eeq 
We see that $\widetilde{S}$ gives a quadratic potential for $\p$, with no dependence on its derivatives. 
From now on we set $\MP = 1$ for simplicity.

Let us consider first the case of $Q_{\mu}$ and $Y_{\mu}$ constant. 
We can make a shift, and center both of them at zero, $Q_{\mu}=0$ and $Y_{\mu}=0$.
The centered action is then:
\beq
\label{actionp}
\S =  \int d^4 x \ \Big[ |\partial_{\mu}   \p|^2 
+   x^{\mu}x_{\mu} |\p|^2  \Big] \ .
\eeq 
The EoM is:
\beq
\label{rhoeq}
\partial^{\mu} \partial_{\mu} \p   -     x^{\mu}x_{\mu}   \p =0 \ .
\eeq
The potential has a positive sign in the time direction, and a negative sign in the space direction.
Signs change exactly like the kinetic term.
This is the same equation of the relativistic harmonic oscillator (see Appendix \ref{apprho}).
We can solve this using the technique of separation of variables.
We make the ansatz
\beq
\p = \p_0(t) \p_s(\vec{x})\ ,
\eeq
and plug it into the  EoM:
\beq
(-\partial_t^2 \p_0 +  t^2 \p_0)   \p_s = \p_0(-\vec{\partial}^2 \p_s + \vec{x}^2 \p_s) \ .
\label{separation}
\eeq
The change of sign in the potential is compensated by the same in the kinetic term.
The solution can be given by the eigenstates of the quantum harmonic oscillator, both for $\p_t$ and $\p_s$, plus the constraint imposed by (\ref{separation}):
\bea
&& \p = \prod_{\rho=0}^3 \frac{1}{\pi^{1/4} \sqrt{2^{k_{\rho}} k_{\rho}!}}  e^{-x_\rho^2 / 2} H_{k_\rho}(x_\rho) \ , \nonumber \\ [1mm]
&&  k_{0} +\frac{1}{2} = \sum_{j=1}^3 {k_j} + \frac{3}{2} \ ,
\label{solreal}
\eea
where $H_k(x)$ are the Hermite polynomials $(-1)^k e^{x^2}(d_x)^k e^{-x^2}$, and $k_0$, $k_i$ are integers.
Here is visible the reason behind the choice of the relative  plus sign between $S$ and $\widetilde{S}$. It is in fact the only way to obtain normalizable solutions to the EoM. Other choices of the relative phase would give inevitably solutions with exponential divergences. 

Since the action is linear in $\p$, we can make a generic superposition of (\ref{solreal}), with different $k$s, 
and still have a solution for the EoM.

Now we discuss  the possibility of  non-centered solutions.
$Q_{\mu}$ is a gauge auxiliary field; it has no kinetic term. The EoM for $Q_{\mu}$ is
\beq
\label{q}
Q_{\mu} = \frac{-i\p^*\partial_{\mu}\p+i\p\,\partial_{\mu}\p^*}{|\p|^2} \ .
\eeq
If $Y_{\mu}$ is constant, we can shift to $Y_{\mu}=0$ and rewrite the action as (\ref{actionp}), with $\nabla_{\mu}$ instead of $\partial_{\mu}$.
$S$ is invariant under all the gauge transformations 
\beq
\label{gauge}
\p \to e^{i\alpha(x)} \p\ , \qquad Q_{\mu} \to Q_{\mu} + \partial_{\mu} \alpha(x) \ ,
\eeq
It is certainly possible to take the solutions (\ref{solreal}) and, making any gauge transformation (\ref{gauge}), obtain new solutions with $\partial Q \neq 0$. But these should not be considered as new solutions. 
We consider solutions (\ref{solreal}), with $Q_{\mu}$ and $Y_{\mu}$ constants, the representative for their gauge equivalent set.
From the EoM (\ref{q}) we can conclude that $\partial_{\mu} Q_{\nu} - \partial_{\nu} Q_{\mu} =0$. In trivial topologies, the solution is always gauge equivalent to a constant.
That proves that if one of the two, $Q_{\mu}$ or $Y_{\mu}$, is constant, then also the other must be (modulo gauge transformations).
Of course the previous argument does not exclude the possibility of non-trivial solutions where both $Q_{\mu}$ and $Y_{\mu}$ are non-constant and non-gauge equivalent to a constant.
But to explore this category we cannot use the trick of rewriting $\S$ as a function of $\p$ only (or $\tp$ only). 



\subsection*{Interactions and Interpretation}

To make an interpretation of these results, we need to introduce interactions, and see how the various linear waves (\ref{solreal}) couple together. The simplest is a quartic term, in both $S$ and $\widetilde{S}$:
\beq
\label{freescalarinteractions}
\S =  \int d^4 x \ \big[(\nabla_{\mu}  \p)^* \nabla^{\mu} \p - (\p^*\p)^2  \big]  +   \int d^4 p \ \big[ (\tnabla_{\mu} \widetilde{\p})^* \tnabla^{\mu} \widetilde{\p} - (\widetilde{\p}^*\widetilde{\p})^2 \big] \ . 
\eeq 
Now that linearity is broken, the waves can be mixed.
There are two limits in which there is a very clear-cut interpretation.
These two limits are dual to each other.

It is now  useful to interpret solutions (\ref{solreal}) as an actual particle that moves in an harmonic oscillator potential. The coordinates of the harmonic oscillator are now $t,\vec{x}$ and we call $T$ the fictitious time. 
The Fourier transform is diagonal in the harmonic oscillator eigenstates
\beq
\tp =  \prod_{\rho=0}^4 (-i)^{k_\rho}  \frac{1}{\pi^{1/4} \sqrt{2^{k_{\rho}} k_{\rho}!}}  e^{-p_\rho^2 / 2} H_{k_\rho}(p_\rho) \ ,
\eeq 
For $k \gg 1$ we can have coherent states. They are beams of minimal dispersion $1$ that oscillates in the potential with $\delta x \sim \delta p \sim \sqrt{k}$. the relation between the $k$s means that the beam oscillates exactly on some light-cone trajectory. 
\begin{figure}[h!t]
\epsfxsize=9cm
\centerline{\epsfbox{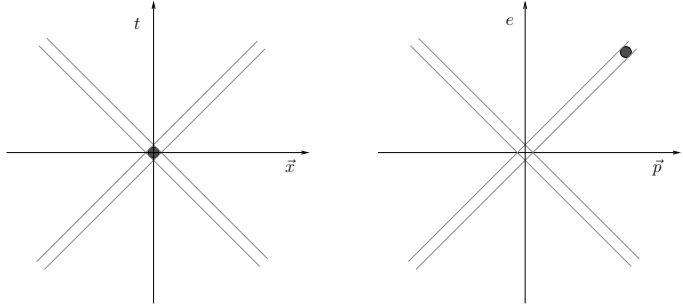}}
\caption{{\footnotesize  A coherent state of the harmonic oscillator. }}
\label{beam}
\end{figure}
The difference between the  $x$ and $p$ oscillations, is just a $\pi/2$ phase shift. So that when one beam is on the center of the potential, the other is at its maximum distance from the center.
One of such configurations is that of Figure \ref{beam},  in which the beam is concentrated at the tip of the space-time light cone.
Its wave function is
\beq
\varphi(x) \simeq  \frac{1}{\pi}  e^{i k_{\mu} x^{\mu}}   \prod_{\rho=0}^4 e^{-x_{\rho}^2/2} \ ,
\eeq
with the relation $k_{\mu}k^{\mu}$  between the momenta.\footnote{Due to the discrepancy in (\ref{solreal}) ($3/2$ on one side, $1/2$ on the other), the coherent state can be a good approximation only for large $k$'s.}

The interpretation in our context is the following. A very small observer would see a world ruled by the action $S$, and the position of the beam in the momentum space, is just the conserved energy-momentum of $S$. 
Waves can interact, scatter with each other, and change their final momentum. 
But always the momentum must belong to the energy-momentum light-cone that is centered in a fixed $Q_{\mu}$. 
That can be understood also from the estimation at the end of Section \ref{dynamics}, where the term $\widetilde{S}$ becomes irrelevant. 
In the other  limit, when the beam is in the center of the energy-momentum light-cone, it is the term $S$ in the action that becomes irrelevant. Now we have 
the same, but dual, interpretation as dynamics in  the energy-momentum manifold.

\subsection*{Classical solution with gravity}

When we add gravity, and its dual counterpart, the structure of interactions is as follows:
\beq
\begin{array}{ccc}
 g  & \   &  \tg\\[1mm]
 \updownarrow & & \updownarrow        \\[1mm]
 \p &  \Leftrightarrow   & \tp
\end{array}
\eeq
The metric $g_{\mu \nu}$ interacts only with $\p$, and the Einstein equations are unchanged, $G_{\mu} = 8 \pi T_{\mu \nu}$.
Where the tensor $T_{\mu \nu}$ is given by
\beq
\int d^4x \sqrt{-g} \  T_{\mu \nu} = \frac{\delta \S}{\delta g^{\mu\nu}} =  \frac{\delta S}{\delta g^{\mu\nu}} \ .
\eeq 
This follows from the fact that the dual-action $\widetilde{S}$ is, by definition, independent on the space-time metric $g_{\mu \nu}$. We then have:
\beq
\label{tensor}
T_{\mu \nu} = -2 (\nabla_{\mu}  \p)^* \nabla_{\nu} \p + g_{\mu\nu} g^{\alpha \beta} (\nabla_{\alpha}  \p)^* \nabla_{\beta} \p \ . 
\eeq
Finally we need to evaluate $T_{\mu\nu}$ on the solutions of the EoM, and here enters the effect of $\widetilde{S}$.
The same is true for the dual-gravity, which is an independent degree of freedom, and satisfies the dual-Einstein equations $\widetilde{G}_{\mu\nu} =8\pi \widetilde{T}_{\mu\nu}$. The tensor $\widetilde{T}_{\mu\nu}$ is obtained from $\delta \widetilde{S}/\widetilde{g}^{\mu\nu}$.


We need to find the formulas to invert the generalized FT (\ref{covfourieruno}), (\ref{covfourierdue}):
\bea
\p(y) &=&   \int \frac{d^4 p \sqrt{-\tg} } {(2 \pi )^{2} } \ \tf_{y}^{-1}(p) \widetilde{\p}(p)\ \ , \\
\widetilde{\p}(q)  &=&   \int \frac{d^4 x \sqrt{-g}} {(2 \pi )^{2} } \ f_{q}^{-1}(x) \p(x)\ \ .
\eea
The $f^{-1}$ are not easy to compute, and we do not know a generic solution for this mathematical problem. Here we just assume their existence.\footnote{Since (\ref{covfourieruno}),  (\ref{covfourierdue}) can be seen as just a linear transformation between functional spaces, the inversion problem is equivalent to finding the inverse of a matrix. 
It can happens that, for certain non-trivial geometries, the map is not injective, and that would require a more careful treatment.}
They are defined in order to make the inversion. In the flat space, the $f^{-1}$ function coincides with the complex conjugate $f^*$, but {\it not} in general when there is curvature.  
The functions $f_q^{-1}(x)$ and $\tf_y^{-1}(p)$ are defined to satisfy
\bea
\label{inv}
 \frac{1}{\sqrt{-g}} \delta(x-x')  &=&   \int \frac{d^4 q  } {(2 \pi )^{4} } \ f_{q}(x) f_{q}^{-1}(x') \ , \\
 \frac{1}{\sqrt{-\tg}} \delta(p-p')  &=&   \int \frac{d^4 y} {(2 \pi )^{4} } \ \tf_{y}(p) \tf_{y}^{-1}(p')  \ .
\eea
In the case of Minkowski space (or gauge equivalent to that), then $f^{-1}$ and $f^*$ are the same, and these are just the usual orthogonality relations between the exponentials. 
For non-trivial metrics that is not true. The fact that $f^*$ does not give the orthogonal function is simple if you think about the interaction between the asymptotic free waves and the metric. Clearly there is some scattering and wave superposition. The mathematical problem is to find an analytic expression for this inverse function.
We can now close the circle and give the relation between $\p(x)$ and $\tp(p)$:
\bea
\label{geninvuno}
\p(x) &=&  \int \frac{d^4 q} {(2 \pi )^{2} }  \frac{d^4 y} {(2 \pi )^{2} }\frac{d^4 p \sqrt{-\tg}} {(2 \pi )^{2} }   \      f_{q}(x) e^{ - i q y  } \tf_{y}^{-1}(p) \ \widetilde{\p}(p) \ , \\
\label{geninvdue}
\widetilde{\p}(p) &=& \int \frac{d^4 y} {(2 \pi )^{2} } \frac{d^4 q} {(2 \pi )^{2} } \frac{d^4 x \sqrt{-g} } {(2 \pi )^{2} }  \       \tf_{y}(p) e^{ i q y  } f_{q}^{-1}(x)   \    \p(x) \ .
\eea

The action $\S$ can be expressed as a function of $\p$ only, even in the presence of gravity
\beq
\label{total}
\S =  \int d^4 x \sqrt{-g}\ \left[ g^{\mu\nu} \partial_{\mu} \p^* \partial_{\nu} \p   + \int d^4 x' \sqrt{-g}\  \p^*(x'){\cal F}(x',x)  \p(x)  \right] \ ,
\eeq 
where ${\cal F}$ is some potential that depends on both $g$ and $\tg$.
In the case of both flat metrics
\beq
{\cal F}(x',x) = \delta(x-x') x_{\mu} x^{\mu} \ ,
\eeq
and we recover (\ref{actionp}).
The EoM  is linear, but non-local
\beq
\partial_{\mu}\big(g^{\mu\nu}  \partial_{\nu} \p(x)\big)  -   \int d^4 x' \sqrt{-g}\  {\cal F}(x ,x')  \p(x')  =0  \ .
\eeq
To compute this correlation-potential ${\cal F}$ we need to use the inversion FT given previously (\ref{geninvuno}) and (\ref{geninvdue}). 
We need also to derive the probe functions $f$ and $\widetilde{f}$.
The solution for ${\cal F}$ is
\bea
{\cal F}(x',x) &=&  \int  \frac{d^4 q'} {(2 \pi )^{2} }   \frac{d^4 y'} {(2 \pi )^{2} }  \frac{d^4 p \sqrt{-\tg} } {(2 \pi )^{4} }   \frac{d^4 y} {(2 \pi )^{2} }   \frac{d^4 q} {(2 \pi )^{2} }   \nonumber \\ [2mm]
&& \Big[   f^{-1}_q(x) e^{iqy}  (\tnabla_{\mu} \tf_{y'}(p))^* 
 \tnabla^{\mu} \tf_y(p) \ 
e^{-iq'y'}\ 
 f_{q'}^{-1 *}(x')\Big] \ .
\eea


\label{cla}


\section{Quantum Aspects}

The classical theory is rather curious. 
It is a realization of the $x \leftrightarrow p$ duality, but in a way opposite to what we would like. 
At high energy scales  $\Lambda \gg  \MP$, it behaves like ordinary theories with action $S$ on the space-time manifold. 
At low energy scales $\Lambda \ll \MP$ is behaves with the dual-action $\widetilde{S}$ on the energy-momentum manifold.
That is clearly phenomenologically bad.

But until now, we have  omitted something important in the discussion: the quantum effects of gravity. 
These can have a huge impact on these ranges of validity.
Unfortunately, at the moment, we do not have quantitative control over these effects. The following discussion, is thus at a very qualitative level.

Now we write the functional integral. 
The first important thing is that we can make an overall integration over $\p$. And this is good both for $\p$ and $\widetilde{\p}$.  So, from this point of view, it is invariant under the duality. In the field theory approach it is equivalent to say that making a path integral over $[d\p]$, or over $[d\widetilde{\p}]$,  is essentially the same thing
\beq
\MP^{-1} \int [d\p] \sim \MP \int [d\widetilde{\p}] \ .
\eeq
The matter fields are the physical objects, the ones that makes $S$ and $\widetilde{S}$ talk to each other.
We then integrate independently over all the possible geometries $g$ and $\widetilde{g}$ of respectively the space-time and the energy-momentum.  We also integrate over the auxiliary gauge bosons $Q_{\mu}(x)$ and $Y_{\mu}(p)$.
\beq
\int     [ d \p]    [dg]     [dQ]     [d\widetilde{g}]     [dY]    \     \exp{(i \S)} \ .
\eeq
The action is completely symmetrical, is just the sum of $S$ and $\widetilde{S}$. 
In order to implement the equivalence principle in both the sides of the duality, we have to use the technique previously introduced in Figure \ref{equivalence}, with the support of the two auxiliary Minkowski spaces $y$ and $q$.

In the Euclidean formulation, field configurations are weighted by $\exp{(-\S_{\rm E})}$. To be properly defined the Euclidean action must be bounded from below, and this forces us to choose a positive sign
\beq
\S_{\rm E} = S_{\rm E} + \widetilde{S}_{\rm E} \ .
\eeq 
The analytic continuation, to be consistent with the Fourier transform must be performed in the following way
\beq
t \to i \tau \ , \qquad e \to  i \epsilon \ .
\eeq
and this is consistent with the fact that the metric must become Euclidean not only in $x_{\mu} x^{\mu}$ and $p_{\mu} p^{\mu}$, but also in the cross-terms $x_{\mu}p^{\mu}$.

And now we come to one of the most important issues, how   $\widetilde{S}$ behaves at low-energy $\Lambda \ll \MP$.
Dual-gravity is here in the strong coupling regime. That follows from the very standard argument. The gravitational part of the action is of the order $ \left( \Lambda \, \delta \tg/ \MP \right)^2 $, where $\delta \tg$ is the fluctuation of the dual-metric.
That means that quantum fluctuation of the dual-metric becomes very big
\beq
\delta \tg \sim \frac{\MP}{\Lambda} \ .
\label{fluc}
\eeq
These fluctuations are just the dual counter part of what happens to the ordinary metric $g$ at high energy scales.\footnote{When fluctuations are non-perturbative, it is expected that changes of topology also occurs in the manifold. 
For this reason this state is also sometimes referred as ``quantum foam.''}
Now we need to consider the dual-matter. The field $\tp(p)$, from (\ref{covfourieruno}), is obtain from a Fourier reconstruction of the waves $\tf_y(p)$, weighted with the function $\p(y)$. The waves $\tf_y(p)$ are subject to the equation (\ref{eomf}) and thus sensitive to the metric fluctuations, especially to the ones that have a similar frequency.
And if the fluctuation are very large (\ref{fluc}), that imply that $\tp(p)$ cannot be reconstructed like the original $\tp(q)$, and is instead scattered all over the energy-momentum manifold.

We certainly need to study this mechanism in more detail, in order to make any progress with this idea of trans-Planckian duality. 

\label{qua}

\section{Conclusion}

In this letter, we presented an idea, quite unusual, about Planck scale physics. 
We quickly summarize the main essence of this proposal.

One of the pillars of quantum mechanics is the wave-particle duality.  
Waves can be expressed, equivalently, as functions over  space-time $x^{\mu}$, or as functions over energy-momentum $p^{\mu}$. 
As long as we do not consider the dynamics, space-time and energy-momentum are completely symmetrical, dual  objects. 
This duality is spoiled by dynamics, and especially by gravity. 
Theories are generally formulated in terms of local Lagrangians in space-time, invariant under   Poincare symmetry. 
Furthermore, after gravity is introduced, space-time becomes a geometric manifold, whose metric is dynamically determined by the energy-momentum tensor.

On the other hand, we have to admit, all the phenomena that we have observed, and that fit into our description, are limited to space-time scales much greater than the Planck length, and energy-momentum scales much smaller than the Planck mass.
So, although it is true that in our theories $x^{\mu}$ and $p^{\mu}$ have different meanings and different roles, it is  also true that we always observe them in very asymmetrical circumstances.
Furthermore, if we try to extrapolate our theory beyond its Planckian boundaries, we inevitably run into trouble (non-renormalizability, strong-coupling). 
The idea we proposed here  is that maybe the asymmetry between $x^{\mu}$ and $p^{\mu}$ that we observe  
is just due to the asymmetric conditions in which we observe the universe. 
We were thus led to search for a theory in which space-time and energy-momentum are completely dual objects, and that reduces to our ordinary description at small energy scales.

There is an immediate consequence, if  this duality were to be realized. 
The energy-momentum is  a manifold itself, with a geometry $\tg_{\mu\nu}$ and the possibility to move around it and have dynamics on it. 
This forced us from the beginning to introduce a $\U(1)$ gauge bundle on both space-time and energy-momentum, with gauge bosons respectively $Q_{\mu}$ and $Y_{\mu}$. 
The expectation value $Q_{\mu}$ contains the information about the point in the energy-momentum manifold in which we are centered right now. 
We then introduced the simplest possible action that takes into account dynamics on the events manifold, and dynamics on the momenta manifold, $\S = S + \widetilde{S}$.
The relative sign was justified by the existence of normalizable solutions to the classical EoM.
Analysis of the classical theory has been performed. It is in fact a realization of the auspicated duality, but high energy phenomena are ruled by the action $S$, and small energy phenomena are ruled by the action $\widetilde{S}$.
Our investigation has been limited to the classical behaviour, although we showed that quantum fluctuations of the metrics $g$ and $\widetilde{g}$ are crucial to determine the regions of dominance.



The existence of asymptotic flat regions, is crucial for the implementation of the whole idea. 
In a cosmological setup, this can be achieved in a FLRW model requiring that the spatial curvature and the fundamental cosmological constant are both zero. 
Both space-time and energy-momentum are asymptotically flat and empty at $t \to \infty$ and $e \to \infty$.
And flatness must apply both to the curvature  of the geometry and to the curvature of the  $\U(1)$ bundle.


We begun in the Introduction with a discussion about the electro-magnetic duality. 
Motivated by this property of gauge fields, we started the search  for an analogous strong-weak coupling duality for the gravitation fields. 
Our conclusion is that gravity itself does not exhibit such property, but with the introduction of an additional structure, the dual-gravity $\widetilde{g}_{\mu\nu}$, it is possible to implement this duality.
One main difference though, is that while electro-magnetic duality exchanges charges without interference with the energy scale, the one described here involves also an exchange between high-energy and low-energy. 
Independently upon the specific model proposed here, we believe that the high-low energy exchange must be a fundamental feature of any attempt to realize the strong-weak coupling duality for gravity.

Our idea  does not put constraints on the matter content or the form of the interactions.
But we have to make a clear distinction between matter field (scalars or fermions), and gauge interactions.
Matter fermions can be treated exactly in the same way we have done for the scalar field $\p$. The self-duality of the Lorentz generators (\ref{generators}), allows in fact for any representation of the Lorentz group to be implemented into this context.
The coupling with the auxiliary bosons $Q_{\mu}$ and $Y_{\mu}$ must be universal, and the same for all matter fields. 
For gauge fields it is instead  different.
If we try to consider a gauge field $A_{\mu}$ as a matter field, and make the Fourier transform of it, we have no chance to make gauge invariance to hold on both sides of the duality.
Gauge fields instead must be treated in the same way as gravity. Both manifolds, space-time and energy-momentum, can have a gauge bundle with some group $G$ and some connection. The two gauge potentials $A_{\mu}$ and $\widetilde{A}_{\mu}$ are two independent degrees of freedom.
Gauge invariance on the two sides of the duality can be obtained with the same trick as Figure \ref{equivalence}. The probe wave functions $f_q(x)$ and $\tf_y(p)$ must transform under the gauge group, like we have done for the $Q_{\mu}$ and $Y_{\mu}$.
It would also be very interesting to try to implement supersymmetry in this context; we do not see any obvious reason that forbids that.


In this short letter we have just presented the idea. 
Clearly we are very far from having control over its quantum aspects and, as we saw, they are a crucial element to decide if this idea can have any relation with the real world. 
\label{con}

\appendix

\section{Relativistic Harmonic Oscillator}
\label{apprho}

We obtained in (\ref{rhoeq}) the equation of a massless relativistic harmonic oscillator (RHO). This object has been considered in the past  as a toy model for relativistic bound states \cite{rho}. 
In those models  $x^{\mu}$ had to be interpreted as the relative coordinate between two elementary particles forming a bound state. 
Our interpretation is different but, since the mathematical problem is the same, is good to discuss the overlap with the already existing literature.

Due to the $x \leftrightarrow p$ duality, the Minkowsky symmetry ${\rm SO}(d,1)$ is enhanced to $\U(d,1)$, and that much in the same way of the Euclidian harmonic oscillator.  The main difference though, is given by the non-compactness of the group, which forces unitary non-trivial representations to be infinite dimensional. The solutions (\ref{solreal}) belongs to a unique, infinite dimensional multiplet. Is in fact clear that the vacuum $ \prod_{\rho=0}^d  e^{-x_\rho^2 / 2}$ is not boost invariant. A representation built above an invariant ground state, such as  $e^{ \pm x_{\rho}x^{\rho}/2}$, has necessarily exponential divergences in time or space. We have not considered the last option in this paper.


\section{Measure Units}
\label{app}

Throughout this paper, we have used natural units in which $\hbar = 1$ and $c =1$. It is instructive to reintroduce them.
In our sub-Planckian universe we have three units of measure: length $[l]$, time $[t]$ and mass $[m]$. 
In the dual, trans-Planckian universe, we have instead: momentum $[\widetilde{l}]=[m l / t]$, energy $[\widetilde{t}]=[m l^2 / t^2]$ and dual-mass $[\widetilde{m}]=[l^2 / t]$. 
The sub-Planckian universe has three fundamental constants: the speed of light $c=[l/t]$, the Planck constant $\hbar=[m l^2 / t]$ and the Newton constant $\GN=[l^3/(m t^2)]$.
The trans-Planckian universe also  has three fundamental constants: $\widetilde{c}=[\widetilde{l}/\widetilde{t}]=[t/l]$, $\widetilde{\hbar}=[\widetilde{m} \widetilde{l}^2 / \widetilde{t}]=[m l^2  /t]$, and $\tGN =[\widetilde{l}^3/(\widetilde{m} \widetilde{t}^2)]=[m t^2 / l^3]$. They are related to the sub-Planckian fundamental constants by:
\beq
\widetilde{c}=\frac{1}{c} \ , \qquad \widetilde{\hbar}= \hbar \ , \qquad \tGN  = \frac{1}{\GN} \ .
\eeq
The Planck constant enters in the wave-length of $e^{ i p^{\mu} x_{\mu} / \hbar } $ and in the Fourier transform. It is self-dual since it is a momentum multiplied by a length. The speed of light enters in
$x^0 =  c t$ and $p^0 = e/c$, and this, in our language, can be interpreted as $\widetilde{c}=1/c$.

\vspace{6pt}

This   work is supported by DOE grant  DE-FG02-94ER40823.

\vspace{6pt}


\end{document}